\documentclass[amsmath,amssymb,aps,prl,twocolumn,showpacs,superscriptaddress]{revtex4-1}
\usepackage{graphicx}
\usepackage{dcolumn}
\usepackage{bm}
\usepackage{xr}
\makeatletter
\usepackage{hyperref}
\usepackage{xcolor}
\usepackage[stable]{footmisc}
\hypersetup{
  colorlinks   = true,
  urlcolor     = black,
  linkcolor    = blue,
  citecolor   = blue
}

\begin{document}

\title{Adhesion as a trigger of droplet polarization in flowing emulsions} 

\author{Iaroslava Golovkova}
\thanks{These two authors contributed equally}
\affiliation{Sorbonne Universit\'e, CNRS, Institut de Biologie Paris-Seine (IBPS), Laboratoire Jean Perrin (LJP), 4 place Jussieu, F-75005 Paris, France.}

\author{Lorraine Montel}
\thanks{These two authors contributed equally}
\affiliation{Sorbonne Universit\'e, CNRS, Institut de Biologie Paris-Seine (IBPS), Laboratoire Jean Perrin (LJP), 4 place Jussieu, F-75005 Paris, France.}

\author{Franck Pan}
\affiliation{Department of Mathematics, Imperial College London, South Kensington Campus, London SW7 2AZ, England, UK.} 

\author{Elie Wandersman}
\affiliation{Sorbonne Universit\'e, CNRS, Institut de Biologie Paris-Seine (IBPS), Laboratoire Jean Perrin (LJP), 4 place Jussieu, F-75005 Paris, France.}

\author{Alexis M. Prevost}
\affiliation{Sorbonne Universit\'e, CNRS, Institut de Biologie Paris-Seine (IBPS), Laboratoire Jean Perrin (LJP), 4 place Jussieu, F-75005 Paris, France.}

\author{Thibault Bertrand}
\affiliation{Department of Mathematics, Imperial College London, South Kensington Campus, London SW7 2AZ, England, UK.} 

\author{L\'ea-Laetitia Pontani}
\email[]{ E-mail: lea-laetitia.pontani@sorbonne-universite.fr}
\affiliation{Sorbonne Universit\'e, CNRS, Institut de Biologie Paris-Seine (IBPS), Laboratoire Jean Perrin (LJP), 4 place Jussieu, F-75005 Paris, France.}

\begin{abstract}
Tissues are subjected to large external forces and undergo global deformations during morphogenesis. We use synthetic analogues of tissues to study the impact of cell-cell adhesion on the response of cohesive cellular assemblies under such stresses. In particular, we use biomimetic emulsions in which the droplets are functionalized in order to exhibit specific droplet-droplet adhesion. We flow these emulsions in microfluidic constrictions and study their response to this forced deformation via confocal microscopy. We find that the distributions of avalanche sizes are conserved between repulsive and adhesive droplets. However, adhesion locally impairs the rupture of droplet-droplet contacts, which in turn pulls on the rearranging droplets. As a result, adhesive droplets are a lot more deformed along the axis of elongation in the constriction. This finding could shed light on the origin of polarization processes during morphogenesis.
\end{abstract}

\maketitle

%
%
%
%
%
%

\section{Introduction}

During morphogenesis, cells both differentiate and self-assemble into tissues and organs with specific forms and functions. For instance, during gastrulation, the Drosophila embryo folds onto itself to produce the ventral furrow that eventually becomes the first tubular shape of the embryo, thus defining the inside-outside geometry of the future organism. This extensive remodeling of tissues is controlled by both biochemical pathways, through soluble morphogens \cite{Kutejova2009a,Meinhardt2009a,Wartlick2009b}, and biomechanical processes, through forces \cite{Mammoto2013b,Heisenberg2013a,Keller2012a} and the regulation of cellular adhesion \cite{Foty2013a,Barone2012}.
The behavior of tissues during morphogenesis is thus strongly determined by their mechanical response, which is controlled by a feedback loop between cellular adhesion and biochemical signaling through the cytoskeleton \cite{Graner1992,Krieg2008,Blanchard2009,Zhang2012b,Farge2011}. Figuring out the properties of the tissue from a materials standpoint is therefore of the utmost importance to fully understand the role of the various processes at play during morphogenesis.

The mechanical properties of tissues and their architecture depend on the properties of the individual cells but also on the adhesion energy between the cells and with the extracellular matrix. As a matter of fact, in the absence of interactions with the extracellular matrix, the level of cell-cell adhesion is directly related to the surface tension of cellular aggregates. As a result it was shown that the level of intercellular adhesion controls the shape and hierarchical organization of cells in aggregates {\it in vitro} \cite{Krieg2008,Schotz2008,Foty1996,Steinberg1994,Krens2011}. These processes were described in the framework of the differential adhesion hypothesis \cite{Steinberg1970}, in which the cohesive cell aggregates are considered as fluids that tend to minimize their interface as a function of the relative strength of cellular adhesion. It was also shown that cell aggregates exhibit mechanical behaviors that depend on the adhesion between cells. For instance, adhesive cell aggregates spread on solid substrates like viscoelastic droplets at short times, but display distinct long time wetting properties when the adhesion is impaired \cite{Douezan2011}. In epithelial monolayers, the correlated rearrangements and cell deformations also indicate that the tissue behaves as a viscoelastic liquid~\cite{Tlili2020}.
Those experimental observations, together with theoretical frameworks \cite{Manning2010}, suggest that soft tissues can be described within a soft matter framework \cite{Gonzalez-Rodriguez2012a}. 

Following this idea, interfacial energy models derived from soap foams were shown to efficiently predict the highly organized cellular structure in organs such as the Drosophila eye \cite{Hilgenfeldt2008}. The behavior of foams, in analogy with tissues, have thus been widely studied under various mechanical constraints. These studies revealed the importance of plastic rearrangements for yielding in those materials \cite{Cantat2013}. Other approaches consist in treating the tissues as fluid-like materials, leading to the modeling of morphogenetic movements based on hydrodynamic theories \cite{Dicko2017,Streichan2018}. Similarly, descriptions borrowed from glassy materials have been recently implemented to describe the collective behavior of cells in developing tissues \cite{Mongera2018}. In this context, the jamming of cells, evidenced by a decrease of fluctuations in the topology of the tissue, directly tunes the material properties of tissues. In turn, it is believed that the jamming transition controls the tissue response to the large stresses during morphogenesis.  Another approach aims to infer the fate of tissues from their static topologies. In this case, the shape of the cells and their packing topology were used to predict the fluidization of tissues \cite{Bi2014,Bi2015,Bi2016, Merkel2018, Merkel2018a,Yan2019}.

\begin{figure*}
	\centering
		\includegraphics[height=7.5cm]{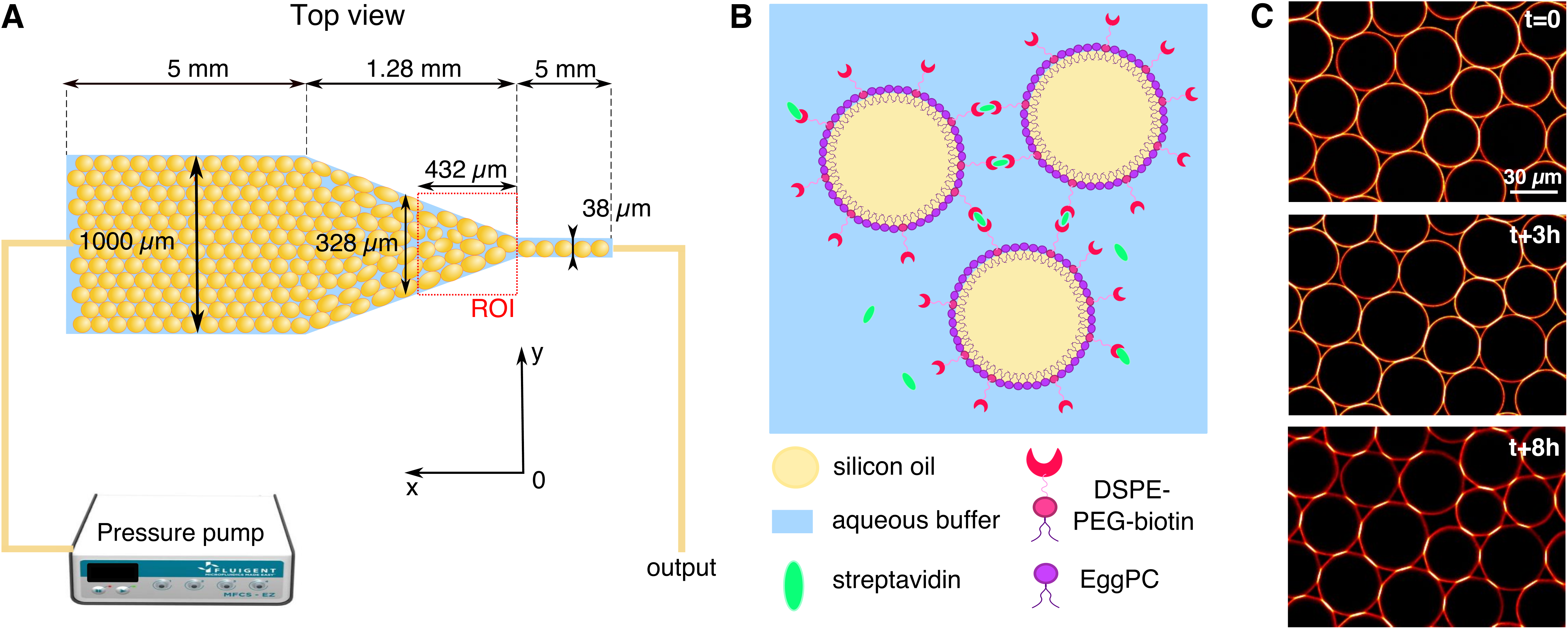}
	\caption{Experimental setup -- (A) The oil in water emulsion is pushed using a pressure pump (P = 15-60 mbar depending on adhesion) through the microfluidic channel that consists of three parts: a 1000~$\mu$m wide channel, a constriction, and 38~$\mu$m wide channel. The depth of the channel is 30~$\mu$m over the whole length, and the diameter of the droplets is $\approx 35~\mu$m. We film the emulsion flow in the area of the constriction situated right before the beginning of the thin channel (see red dashed square for region of interest). (B) Schematic representation of biotin-streptavidin-biotin bonds forming between the contacting surfaces of the droplets stabilized with phospholipids. (C) Progressive formation of adhesive patches over time. The top confocal image shows that Alexa-555 streptavidin fluorescence is more homogeneously distributed over the surface of the droplets at the beginning of the experiment. Over time, biotin-streptavidin-biotin bonds form at the droplet-droplet contacts (middle image) until they enrich into clear adhesive patches with an increased fluorescence signal, allowing to isolate them through image analysis. Note that the formation of the patches depletes the fluorescence level on the free edge of the droplets making it appear more red over time. }
	\label{Fig1}
\end{figure*}

Here, we propose to bridge the gap between biological systems and soft matter frameworks by using biomimetic emulsions to decipher the collective dynamics and material properties of tissues during remodeling. In particular, we address the impact of cell-cell adhesion on the mechanical properties of tissues by using functionalized adhesive emulsions. In our previous work, we showed that weakly attractive droplets displayed impaired plastic rearrangements under flow \cite{Golovkova2020}. Here, we propose to directly mimic intercellular adhesion by introducing specific interactions between the droplets \cite{Pontani2012,Feng2013,Pontani2016}. Such biomimetic systems have already been shown to reproduce the minimal adhesive and mechanical properties of tissues in static experiments \cite{Pontani2012}. Their specific interactions are here introduced through biotin-streptavidin-biotin bonds that are allowed to form between the surfaces of contacting droplets. The energy of those binders is comparable to the one of cadherin homophilic interactions in tissues \cite{Katsamba2009}. 
Moreover, the fluidity of the droplets surface allows the binders to diffuse on the droplets surface and to aggregate into adhesion patches at each droplet-droplet contact. At equilibrium, the size of the patch can be roughly determined by the balance between the gain in adhesion energy and the loss in elastic energy due to the flattening of the droplet surface in the patch \cite{Pontani2012}. 

We study the response of these systems under mechanical stress. In order to impose a mechanical perturbation on the assembly of adhered droplets, we push them through a 2D microfluidic constriction (see Fig.\,\ref{Fig1}A). This geometry forces rearrangements in the emulsions, allowing us to study their elasto-plastic response, but also aims to mimic the convergent extension of epithelial tissues that is essential during embryogenesis \cite{Sutherland2020}.
We find that adhesion does not affect the rearrangements topology and that the size of avalanches exhibit the same statistics for all experimental conditions. This observation is further confirmed in simulations that allow us to explore different droplet size polydispersities, deformabilities and adhesion energies. These simulations similarly evidence avalanche size statistics to be independent of adhesion. However, when exploring experimentally the individual T1 events, we find that the local dynamics are slowed down in adhesive emulsions as the binding patches prevent droplet-droplet separation during rearrangements. In turn, we observe that adhesive patches lead to large scale deformations across all droplets in the constriction. In addition to being more deformed, we find that the droplets are also more aligned with each other, which could be the signature of an adhesion-induced polarization process in elongating tissues.

\section{Materials and Methods}

\subsection{Emulsion preparation}

Oil in water emulsions were prepared using a pressure emulsifier, as described in \cite{Golovkova2020}. After emulsification, the oil droplets were stabilized with phospholipids in order to make adhesive biomimetic emulsions, as shown in Fig.\ref{Fig1}B. Firstly, 9 mg of egg L-$~\alpha$-phosphatidylcholine (EPC) lipids and 1mg of DSPE-PEG(2000) biotinylated lipids (Avanti Polar Lipids) were dried under nitrogen and dissolved in 500$~\mu \mathrm{L}$ of dimethyl sulfoxide (DMSO, from Sigma Aldrich). This mixture was then added to 5 mL of high SDS aqueous buffer (5mM SDS, 10mM Tris, pH=7.5). The resulting solution was sonicated for 30 minutes. We then added 2mL of emulsion cream to the phospholipid containing buffer and left to incubate overnight at 4$~^\circ \mathrm{C}$. After incubation the emulsion was washed with 100 mL of high SDS buffer (5mM SDS, 10mM Tris) in a separating funnel and set for a second round of stabilization. During this second round, the SDS concentration in the aqueous phase is lowered in order to favor the repartition of lipids at the droplets surface, instead of SDS, while still keeping them from coalescing. The low SDS aqueous solution used for this last round therefore contains 1mM (instead of 5mM) SDS while the rest of the procedure for lipid dissolution remains unchanged. After this last incubation, the emulsion is washed again in 100 mL of low SDS buffer (1mM SDS, 10mM Tris) in a separating funnel. The resulting droplets display an average diameter of 35$~\mu \mathrm{m}$ (with a size polydispersity of 21$\%$) and are stable over several weeks at 4$~^\circ \mathrm{C}$.

Before running the experiments, the droplets are functionalized with streptavidin Alexa Fluor 555-conjugated (Invitrogen). To this end, 200$~\mu \mathrm{L}$ of the emulsion cream is mixed with 3.6$~\mu \mathrm{L}$ of streptavidin (1 mg/mL) and 200$~\mu \mathrm{L}$ of low SDS buffer. The resulting solution was incubated for 1 hour at room temperature to allow the streptavidin to bind to the biotinylated lipids on the surface of the droplets. The emulsion was then washed twice in the 800$~\mu \mathrm{L}$ of low SDS buffer and once with 1 mL of a water/glycerol mixture (60:40 v:v) containing 1mM SDS, 10mM Tris, 10mM NaCl and 0.05 mg/mL casein ($~\beta$-casein from bovine milk, Sigma Aldrich). The water/glycerol mixture ensures that the optical index of the continuous phase matches better the one of the oil droplets for transparency, while salt favors adhesion by dampening electrostatic repulsion between the droplets \cite{Pontani2012}.

\subsection{Experimental set-up}

The microfluidic channels are engineered following the techniques described in \cite{Golovkova2020}. The channel consists of three sections: at the inlet the channel is first 1 mm wide over 5 mm length, then the width is reduced from 1 mm to 38$~\mu \mathrm{m}$ over a length of 5 mm, and then the channel remains 38 $~\mu \mathrm{m}$ wide over 5 mm before the outlet (see Fig.\ref{Fig1}A). In order to maintain the droplets in a monolayer, the depth of the setup is adjusted to 30$~\mu \mathrm{m}$, thus facilitating image analysis.

\begin{figure}
	\centering
		\includegraphics[width=9cm]{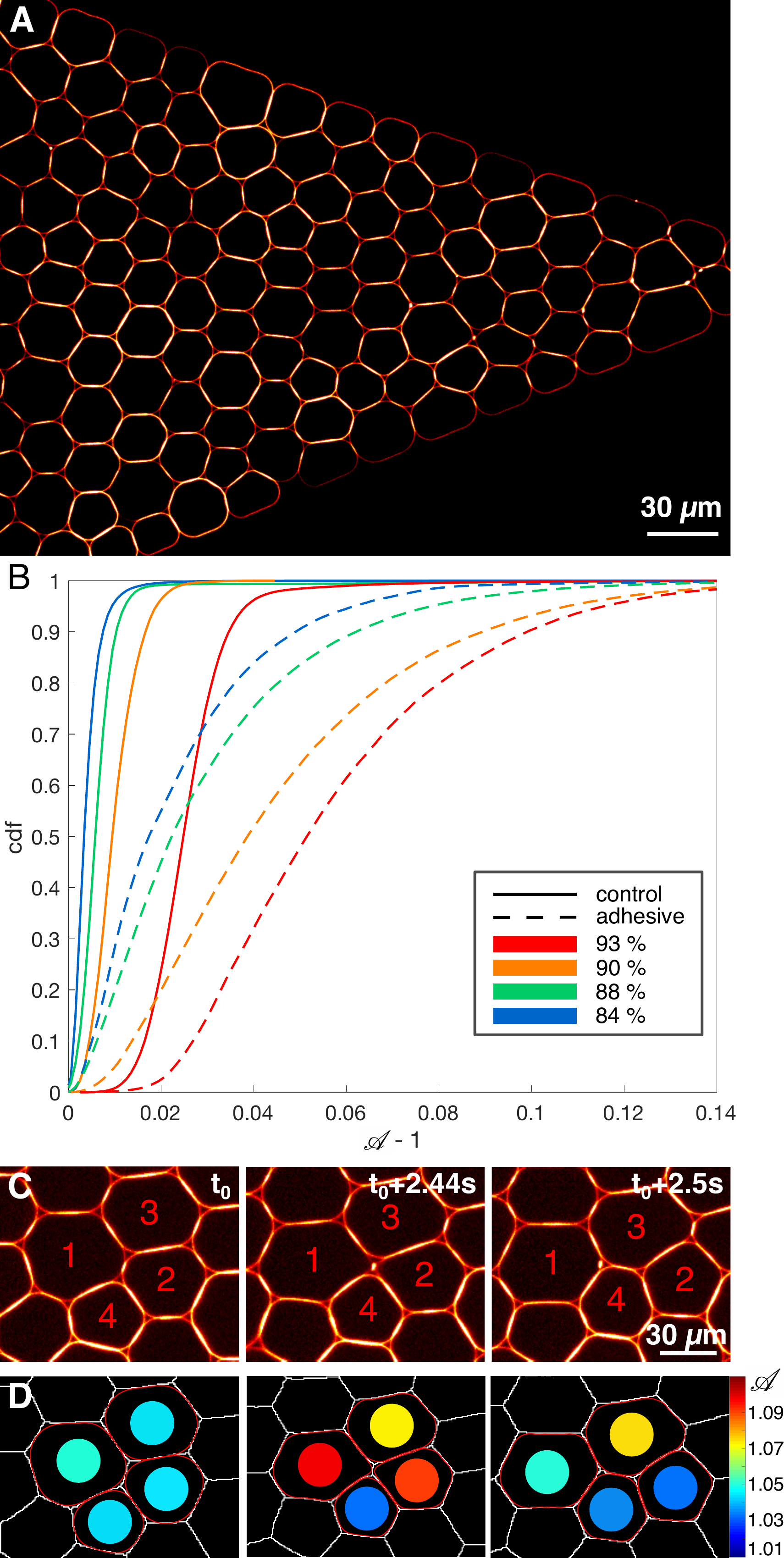}
	\caption{ Confocal imaging and analysis -- (A) Confocal image of an adhesive emulsion in the constriction. (B) Cumulative distributions of deformation $\mathcal{A}$ - 1 for all droplets in the region of interest for adhesive (dashed lines) and non-adhesive control (solid lines) emulsions across packing fractions ranging from $\phi_l=84$ to $93 \%$. (C) Confocal images of four adhesive droplets undergoing a T1 event. Droplets 1 and 2 are first connected through an adhesive patch (left panel), are then pulled apart (middle panel) and are not neighbors anymore (right panel). In the meantime droplets 3 and 4 gain a contact at the end of the event. (D) Result of the image analysis performed on (C). Voronoi cells are drawn in white lines, droplet contours are shown in red and the color of the disc inside each droplet codes for its deformation $\mathcal{A}$.}
	\label{Fig2}
\end{figure}

After passivating the channel with a solution of 0.25 mg/mL casein for 40 minutes, the emulsion is flowed through the channel using a pressure pump (MFCS-8C Fluigent). After the droplets fill the channel, the pressure is decreased to stop the emulsion flow (P = 5 mbar) and the droplets are left overnight to allow the droplets to pack and the adhesion patches to grow (see Fig.\ref{Fig1}C). After the incubation phase, the emulsion is flowed in the channel under constant pressure (P = 15-60 mbar depending on adhesion) and imaged in the constriction area through confocal microscopy with a 20x objective (exposure time = 20 ms, frame rate = 15 fps, see Fig.\ref{Fig2}A).

\subsection{Numerical Simulations}

In order to explore a wide range of parameters, we develop a computational model for adhesive emulsions that is based on the deformable particle model (DPM) recently introduced by Boromand et al. \citep{Boromand2018,Boromand2019}. In the case of 2D emulsions, particles deform in response to mechanical stresses to minimize their perimeter while keeping their area fixed. Modeling each of the $N$ emulsion droplets as a deformable polygon with $N_v$ circulo-line edges with width $\delta$, our model relies on the minimization of the following potential energy: 
\begin{equation}
U_{DP} = \gamma \sum_{m=1}^N \sum_{i=1}^{N_v} l_{m,i} + \frac{k}{2} \sum_{m=1}^N (a_m - a_{m0})^2 + U_{int}
\end{equation}
where $l_{m,i}$ is the length of the circulo-line between vertices $i$ and $i+1$ and $a_m$ is the area of droplet $m$. The first term in $U_{DP}$ is proportional to the perimeter of the droplet with a proportionality constant equal to a line tension $\gamma$. The second term is a penalization term quadratic in the distance between the area of the droplet and a target area $a_{m0}$ with compressibility coefficient $k$.

Finally, $U_{int}$ represents the interaction potential energy between two droplets; it is composed of a repulsive term and an attractive term (see details in SI\dag). Upon contact, overlaps between interacting droplets are penalized by introducing a purely repulsive interaction potential $U_{r}$ between all pairs of circulo-lines of different droplets. Upon contact, two droplets are also subjected to contact mediated adhesion. For instance, after the initial contact between a vertex of droplet $i$ and a vertex or an edge of droplet $j$ is made, droplets $i$ and $j$ are subject to an attractive force derived from the interaction potential $U_a$ \cite{Chaudhuri2012}. See supplementary information\dag~for the detailed expressions of the repulsive and attractive energy terms.

We use the same constriction angle as in the experiments (see movie\dag). We flow the emulsion through the constriction by subjecting each particle vertex to a constant force in the direction of the channel. We use periodic boundary conditions along the axis of the microfluidic channel, i.e. that droplets exiting the constriction re-enter the channel ahead of the constriction. Provided these forces, we integrate the equations of motion for the vertices in the overdamped limit. 

We perform simulations with $N=128$ deformable droplets with $N_v = 16$ vertices per droplet. We vary the line tension $\gamma$ and the adhesion strength $k_a$ keeping all other parameters fixed for monodisperse emulsions and polydisperse emulsions with $20\%$ polydispersity (see in SI\dag). We place ourselves in the limit of non-overlapping, nearly incompressible emulsion droplets.

\subsection{Data Analysis} 

\subsubsection{Tesselation and tracking} 

Raw images are segmented using Ilastik \cite{Berg}. The segmented images are then skeletonized and droplets are detected using Fiji. Droplets, as well as channel boundaries, are then indexed directly on the segmented image. A surface Voronoi tessellation is finally performed on these processed images to identify the Voronoi cells corresponding to each droplet.

A table of neighboring relationships between droplets and Voronoi cells is generated using the Region Adjacency Graphs from the Python Scikit-image package~\cite{van2014scikit}. We then obtain the list of neighbors at each time for each droplet in the constriction and measure the size of droplet-droplet contacts as well as the length of the edge between neighboring Voronoi cells. The droplets are tracked with a custom Python tracking algorithm allowing us to compute instant velocities of droplets and Voronoi cells.

\subsubsection{Deformation} 

We measure droplet deformations following the method used in \cite{Golovkova2020}. To avoid artificial measurement noise due to finite image resolution, we fit successions of osculating arcs of circles around the droplet contours. The computed shape parameter $\mathcal{A} = p^2/4 \pi a$, with $p$ the perimeter and $a$ the surface of the identified droplet, and local packing fraction $\phi_l$ are then calculated from this fitted contour, as shown in Fig.\ref{Fig2}D. Note that we exclude the droplets whose corresponding Voronoi cells touch the walls of the channel. In parallel, we also fit each droplet with an ellipse, and use its aspect ratio and orientation of the major axis to study elongation and alignment of the droplets in the constriction. Further details of the image analysis can be found in SI~\dag.

\subsubsection{T1 events detection}

By tracking droplets and their neighborhood over time, we identify the formation or rupture of droplet-droplet contacts and edges between Voronoi cells. This allows us to identify individual T1 events by considering the neighborhood of droplet quadruplets as shown in Fig.\ref{Fig2}C. We then examine avalanche phenomena by considering T1 events that occur during a given time window and that are connected by neighboring droplets.

To do so we define an adimensional time $t= t^{*} \langle V \rangle / \langle R \rangle$ with $t^{*}$ the elapsed time in seconds, $\langle V \rangle$ the mean flow velocity and $\langle R \rangle$ the mean radius of droplets that are both averaged over all droplets in all frames of each movie.
T1 events whose cells were neighbors to each other within a specified time window (here, 0.4 in adimensional time) are grouped in a common avalanche event (see SI\dag). For simulation data, the T1  were similarly identified from the loss and gain of physical contact between quadruplets of droplets, and grouped in avalanches using the same adimensional time window. We quantify avalanche sizes by measuring the total number of droplets participating in the same avalanche.

During a rearrangement, we also measure the speed at which contacts between voronoi cells are shrinking before the actual neighbor exchange. To do so, we measure $\frac{\Delta l_e} {\Delta t} / {\langle V \rangle } $, where $l_e$ is the contact length between neighboring Voronoi cells, $\Delta l_e = l_e(\text{frame } n) - l_e(\text{frame }n+1)$ and $\Delta t$ the time between two consecutive frames. We measure it for all the droplets involved in a T1 event during the adimensional time window $\left[t_0-10 , t_0\right]$, $t_0$ being the exact moment of neighbor exchange.

\section{Results}

We inject and flow emulsions in a microfluidic constriction as depicted in Fig.\ref{Fig1}A (see also supplementary movie~\dag). In the large channel the emulsions spans about 30 droplet diameters, and it progressively reduces to one droplet in the thin channel. We experimentally tested both non-adhesive and adhesive emulsions (see Materials and Methods). When the droplets are adhesive one expects their flow to be hindered and the emulsion to behave more elastically, whereas an assembly of repulsive droplets, for which rearrangements can be performed at lower energetic cost, should be more plastic. Indeed, when measuring the shape parameter $\mathcal{A}$ of all droplets in the constriction, we find that they are globally more deformed in the case of adhesive emulsions for all packing fractions (see Fig.\ref{Fig2}B), which is consistent with previous work~\cite{Golovkova2020}.

In agreement with this global observation, much higher pressures need to be applied for adhesive emulsions to flow in the constriction compared to non-adhesive ones. On average, in our experimental conditions, one needs to apply about 15-20 mbar with the pressure controller for repulsive droplets, as opposed to 30 to 60 mbar for adhesive ones. At the macroscopic scale these two systems therefore exhibit very distinct material properties. We explore in what follows the microscopic origin of this difference in behavior.

\subsection{Topology and local dynamics of rearrangements}

\begin{figure}
	\centering
		\includegraphics[width=9cm]{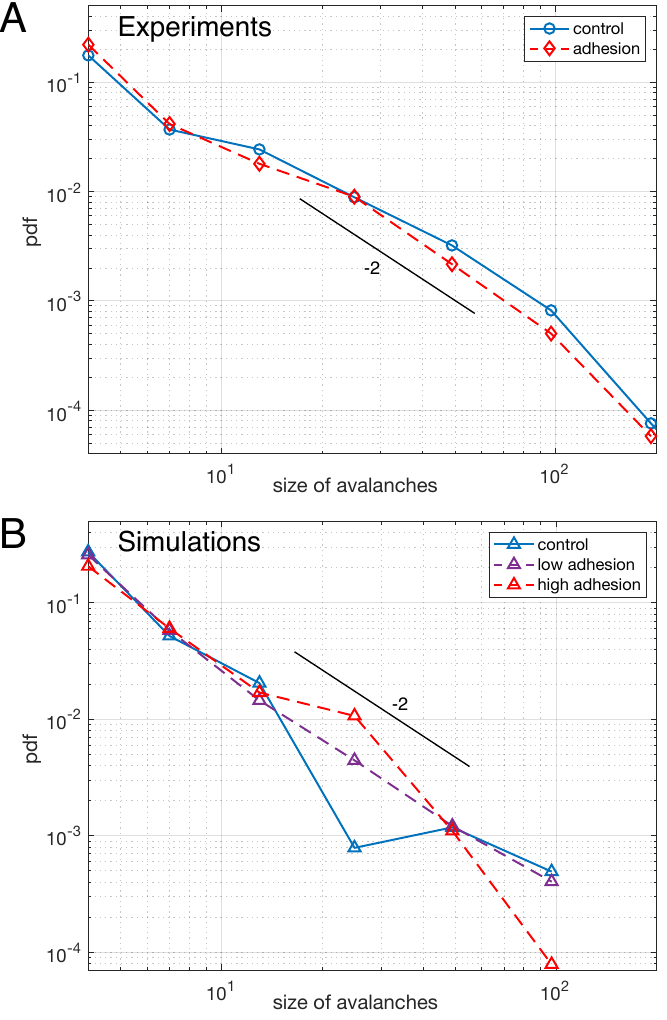}
	\caption{Avalanche statistics for experimental data (A) and numerical simulations (B) - (A) Distributions of avalanche size for adhesive (red dashed line) and non-adhesive (blue solid line) emulsions cannot be distinguished. (B) Distributions of avalanche size for polydisperse packings of highly deformable droplets (lowest $\gamma$) without adhesion (blue solid line), with low adhesion (purple dashed line) and high adhesion (red dashed line), see SI\dag for values of $k_a$ and $\gamma$. All curves are averaged over 5 repeats of simulations performed with the same parameters. The distributions are not significantly different between each other. The logarithmic binning as powers of two is used for the x-axis in both panels. The maximum cluster size that we measure corresponds to an avalanche over the entire field of view, indicating that the choice of time window does not artificially exclude large avalanches from the analysis. In both panels a line corresponding to a power law with exponent $-2$ was drawn as a guide to the eye.}
	\label{Fig3}
\end{figure}

We first study the properties of these two different kinds of emulsions by examining the topology of droplet rearrangements such as T1 events depicted in Fig.\ref{Fig2}C-D. Indeed, it was previously shown that the rearrangements of monodisperse droplets are correlated and ordered in space and time when going through a constriction \cite{Gai2016,Golovkova2020}. In particular, T1 events are aligned along disclination planes that are regularly spaced. Here, we do not expect to see such patterns emerge in the constriction, even in the absence of adhesion, as our droplets exhibit a 21$\%$ size polydispersity. Instead, our experiments display a spatially heterogeneous and intermittent flow which is commonly observed in nature during avalanching. Indeed, a large variety of physical systems \cite{Lebouil2014,Dalton2001,Hartley2003,Baldassarri2006,Amon2012,Denisov2017,Maloney2004,Maloney2009,Otsuki2014,Dahmen2011,Lin2014} generically exhibit intermittent dynamics which is characterized by a slow build-up and a rapid release of stress in the system when subjected to a slow continuous loading. Few studies have looked at this intermittent dynamics at the local scale \cite{Howell1999,Kabla2007,Daniels2008,Lebouil2014,Bares2017}. Here, having access to the whole dynamics during the emulsion flow, we examine the statistics of avalanche sizes through a measure of the local plastic rearrangements. 

More specifically, we measure the size of the avalanches for different adhesion conditions. We define the avalanche size as the number of droplets participating in spatially and temporally connected rearrangements during a given time window. In particular, T1 events whose cells are neighbors at any point within the time window are grouped in the same avalanche. Under all experimental conditions, the avalanches sizes are distributed according to power-law distributions and are surprisingly indistinguishable with or without adhesion, as shown in Fig.\ref{Fig3}A. Although one would expect adhesion to give rise to longer range effects, large avalanches do not seem to prevail in adhesive emulsions. Moreover, in both conditions the distribution of avalanche sizes reasonably follows a power law with a $-2$ exponent, in agreement with previous results obtained for 2D granular packings~\cite{Bares2017}.

To confirm this observation, we performed numerical simulations (see Materials and Methods), allowing us to systematically vary the adhesion energy, droplet deformability and polydispersity. We first examined packings with the same polydispersity as in our biomimetic emulsions. We find that adhesion does not affect significantly the distribution of avalanche sizes as shown in Fig.\ref{Fig3}B, which confirms our experimental findings. However, in the lowest deformability condition, a difference between monodisperse and polydisperse packings clearly exists (see SI\dag). Indeed, our results show that monodisperse packings exhibit an excess of large avalanches for all adhesion energies. This result is consistent with the idea that low deformability monodisperse particles exhibit a higher crystalline order leading to large rearrangements taking place along disclination planes~\cite{Zaiser2006, Gai2016}.

\begin{figure}
	\centering
		\includegraphics[width=9cm]{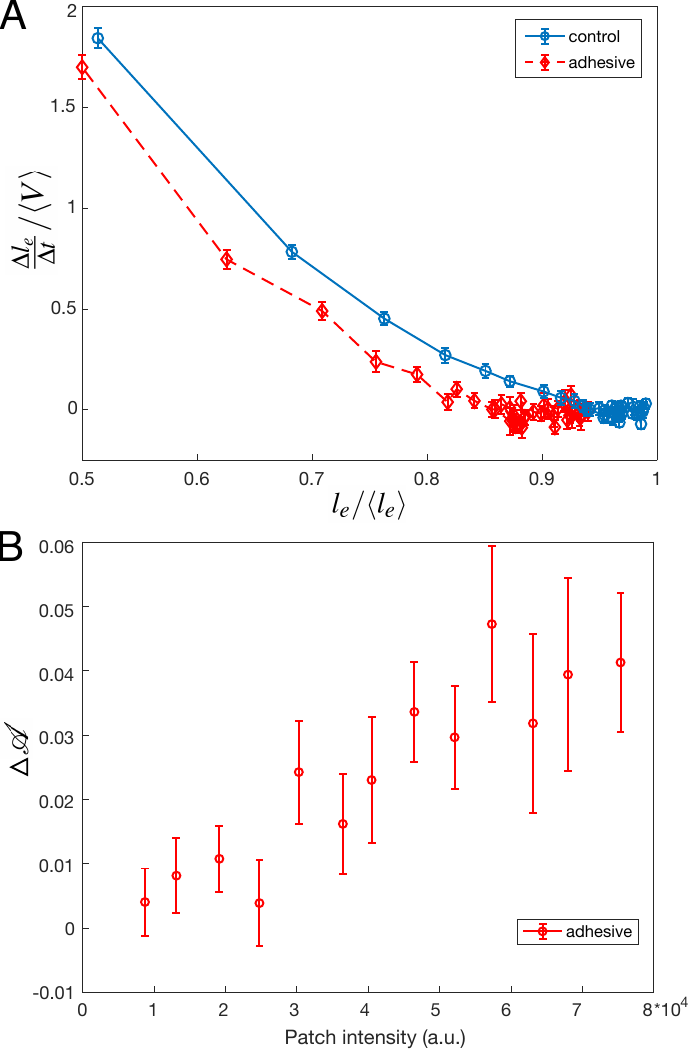}
	\caption{Properties of individual rearrangements - (A) Rate of change of shrinking for dislocating edges of length $l_e$ normalized by the average velocity of the flow $\langle V \rangle$ plotted as a function of $l_e$ normalized by the global average edges length $\langle {l_e} \rangle$. For a given $l_e$, a dislocating edge length is always disappearing more slowly for adhesive emulsions. (B) Difference in droplet deformation $\Delta \mathcal{A}$ before and after they undergo a T1 event as a function of adhesive patch intensity. An increase of patch intensity, i.e. an increase in binding energy, corresponds to higher values of droplet deformation accumulated before a T1 event.}
	\label{Fig4}
\end{figure}

While we could not find evidence of the effect of adhesion on the statistics of avalanches sizes, the consequences of adhesion can be evidenced locally by examining the dynamics of individual rearrangements. To do so, we have first measured the speed at which the length of the dislocating edge decreases during a T1 event. We find that the edge length shrinks more slowly for adhesive droplets for all edge lengths, as shown in Fig.\ref{Fig4}A. Here, the presence of adhesion helps stabilize short edges and slows down the dislocation process by adding a strong energetic barrier to the formation of the rosette preceding the actual neighbor exchange \cite{Bi2019}. However, once the contacting droplets have been separated, the growth of a new edge takes place faster for adhesive droplets (data not shown). This is due to the additional accumulated pressure necessary to break the adhesive contact, which pushes the new droplets in contact more promptly.

In conclusion, avalanche size statistics in flowing emulsions is not affected by adhesion. In fact, the signature of adhesion only lies in the local dynamics of T1 events rather than in long range collective effects. We next study the impact of these local dynamics on droplet deformations.

\begin{figure*}
	\centering
		\includegraphics[height=5.5cm]{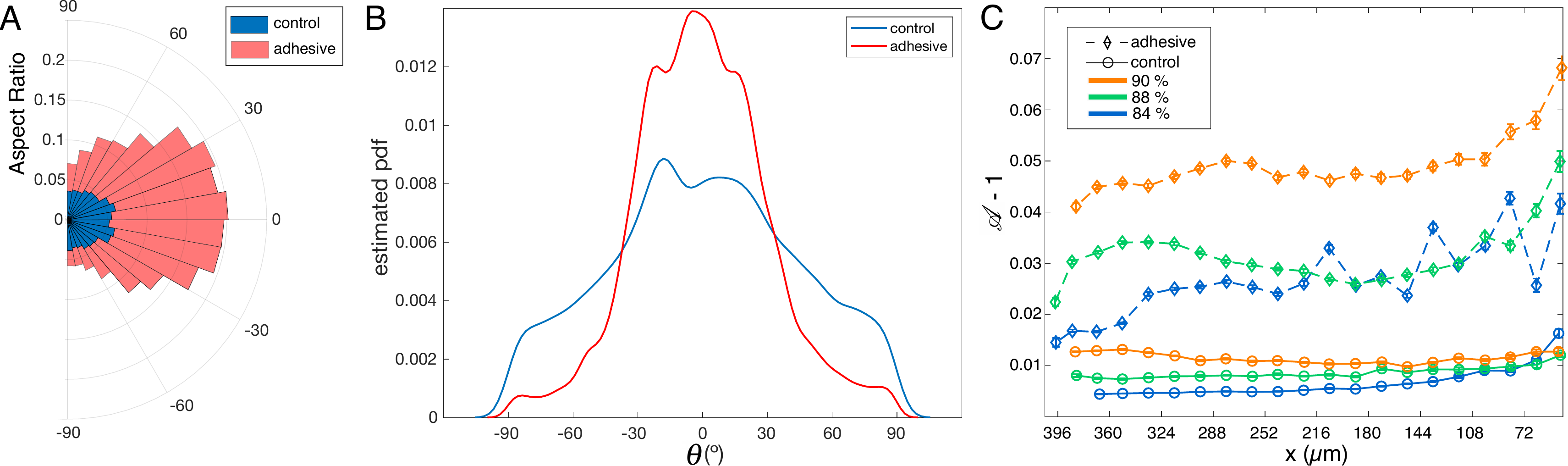}
	\caption{{Analysis of droplet deformation} --- {(A) and (B)} Analysis of the ellipses fitted to control (blue) and adhesive (pink) droplets in packings with average $\phi_l = 88\%$. {(A)} Aspect ratios as a function of ellipse orientation. The aspect ratios for adhesive droplets are significantly larger for all considered angles. {(B)} Distributions of ellipse orientations $\theta$ with respect to the x axis in the constriction. Adhesive emulsions yield a narrower distribution than control ones. {(C)} Average deformation $\mathcal{A}$-1 of the droplets along the x axis of the channel for the adhesive (diamonds) and control (circles) emulsions at various packing fractions. The deformation is averaged in 35 $\mu$m bins along the x axis, x=0 corresponding to the entry of the thin channel. The error bars correspond to the standard error of mean for the distribution of $\mathcal{A}$ obtained in each bin.}
	\label{Fig5}
\end{figure*}

\subsection{Droplet deformations}

As a consequence of impaired rearrangements, the droplets should be more deformed during T1 events in adhesive emulsions. Indeed, the adhesion patches induce pulling forces on the droplets in addition to the compressive forces induced by the constriction geometry. To relate the locally slowed down rearrangements to increased deformations, we have examined the deformation of droplets involved in T1 events by measuring their shape parameter $\mathcal{A} = p^2/4 \pi a$ over the course of the rearrangement. In particular, we quantify the difference between the level of deformation before and after a T1 by measuring $\Delta \mathcal{A} = \frac{\mathcal{A}_1^{t^-} + \mathcal{A}_2^{t^-}} {2} - \frac{\mathcal{A}_3^{t^+} + \mathcal{A}_4^{t^+}} {2}$, where $\mathcal{A}_1, \mathcal{A}_2 $ are droplets that were in contact before a T1 event, and $\mathcal{A}_3, \mathcal{A}_4$ are droplets that became in contact after the T1 event. $t^-$ and $t^+$ are the frames just before and after the rearrangement respectively. For non-adhesive emulsions, we find that the distribution of $ \Delta \mathcal{A}$ is symmetric around zero, indicating that droplet deformations are identical before and after the rearrangements (see SI\dag). In contrast, we find that this distribution becomes asymmetric when droplets interact through specific binders. However, after detachment, the droplets do not exhibit any excess in deformation and thus behave like repulsive droplets. This makes sense because in our system adhesion is short range and dense adhesive patches form on the timescale of hours. 

Considering this scenario, we finally relate the amount of excess deformation $\Delta \mathcal{A}$ during rearrangements to the binding energy between droplets. To do so, we plot $\Delta \mathcal{A}$ as a function of the streptavidin fluorescence intensity at the dislocating contact area and find that a higher intensity, meaning a higher binding energy, directly correlates with droplets that are more deformed prior to rearrangements (Fig.\ref{Fig4}B). This directly links deformation levels in the emulsion to binding energy between droplets.

To track the global effect of this local excess of deformation during rearrangements, we have fitted ellipses to all droplets in the field of view and measured the aspect ratio of the ellipses as well as the orientation of their major axis with respect to the horizontal x axis defined in Fig.\ref{Fig1}A. As shown in the polar plot in Fig.\ref{Fig5}A, the aspect ratio of the droplets is significantly higher for adhesive emulsions for the whole range of orientations. Moreover, the distribution of ellipse orientations is more peaked in the case of adhesive emulsions (Fig.\ref{Fig5}B), while control droplets tend to follow the orientation of the flow field that is measured by tracer particles in the constriction (see SI~\dag). 

In order to quantify droplets deformation independently of their orientation, we measure the shape parameter $\mathcal{A}$ for all droplets in all experiments. When plotted along the x axis, we observe that the shape parameter of adhesive droplets is much higher for all considered packing fractions (see Fig. \ref{Fig5}C). In addition, this high deformation does not relax back to the values measured for non adhesive emulsions even far from the outlet (i.e. $\approx$10 droplet diameters away from the entry of the small channel). This indicates that the effects of adhesion on droplet deformations are long-ranged, which suggests that forces are transmitted more efficiently through the emulsion in the presence of adhesive patches.

\section{Conclusion}

Intuitively, cell-cell adhesion is expected to rigidify biological tissues, providing them with an elastic response to an applied force. In cellular aggregates, the level of cadherin expression has indeed been shown to control the wetting properties on 2D surfaces  \cite{Douezan2011}, while in developing tissues loss of cadherin function can induce a lowering of the yield stress \cite{Mongera2018}. This effect of adhesion on the bulk material properties of tissues is also observed indirectly in our biomimetic emulsions. Indeed, a much higher pressure is needed to induce flow in the case of adhesive emulsions. However, passed that threshold force, both repulsive and adhesive emulsions can flow and go through a constriction. We observed that the flow of emulsions under continuous load exhibits a spatially and temporally heterogeneous dynamics that are characteristic of yield-stress materials. Furthermore, our experiments and simulations both show that the avalanche size statistics is independent of adhesion but weakly dependent on the presence of crystalline order in the spatial structure of the emulsions. 
 
However, the way those rearrangements take place differs with or without adhesion. Indeed, adhesion prevents the detachment of bound droplets, leading to slowed down dynamics prior to the first droplet-droplet contact loss in T1 events. As a result, droplets exhibit larger deformations and they tend to align with the direction of tissue elongation. These long-range cell elongations could be the onset of symmetry breaking in tissues, thus inducing signaling pathways during morphogenesis. Indeed, tissue shape changes can be due to a combination of external forces as well as intrinsic forces. This is the case for the process of convergent extension that is very conserved across metazoans \cite{Sutherland2020}. These intrinsic forces usually emerge from an anisotropy of contractility in individual cells \cite{Martin2009} and recent studies highlighted the importance of in-plane anisotropy \cite{Chanet2017, Doubrovinski2018}. In this case, the cytoskeleton is remodeled and acto-myosin contractility can be increased at cell-cell junctions that are perpendicular to the extension axis \cite{Martin2020, Heer2017}. Interestingly, a recent study also evidenced the importance of cell alignment to predict the fate of tissues and highlighted its impact for rapid morphogenetic movements such as the convergent extension of the drosophila germband \cite{Wang2020}.

In this context, our results suggest that adhesion could participate to morphogenetic processes by inherently making the cells anisotropic when tissues start to be elongated, thus providing a positive feedback loop between external forces and the intracellular response. Beyond these findings, our biomimetic approach paves the path to unraveling other biological mechanisms in the future, such as the role of the extracellular matrix or that of differential adhesion during morphogenetic processes.

\section*{Conflicts of interest}
There are no conflicts to declare.

\section*{Acknowledgements}

The authors thank Gladys Massiera and Laura Casanellas for fruitful discussions, as well as Jacques Fattaccioli for letting us use his pressure emulsifier. L.-L. Pontani also acknowledges financial support from Agence Nationale de la Recherche (BOAT, ANR-17-CE30-0001) as well as financial support from Emergence(s) Ville de Paris.

%

\end{document}